\documentclass[usegraphicx,usenatbib,usedcolumn,useAMS]{mn2e}

\usepackage{times}

\title[Galaxy Zoo: unusual clusters]%
{Galaxy Zoo: an unusual new class of galaxy cluster}
\author[M. F. Pedbost et al.]{%
  Marven F. Pedbost,$^{1}$\thanks{E-mail: team@galaxyzoo.org}
  Trillean Pomalgu,$^{1}$
  and the Galaxy Zoo team
\smallskip\\
  $^{1}$Institute of Cosmology, University of Brentwood,
    Brentwood, Essex, HG4 2TG, UK.
    \vspace*{0.8cm}
}

\begin{document}
  
\date{}

\pagerange{\pageref{firstpage}--\pageref{lastpage}} \pubyear{2009}

\maketitle

\label{firstpage}

\begin{abstract}
We have identified a new class of galaxy cluster using data from the Galaxy Zoo project.  These clusters are rare, and thus have apparently gone unnoticed before, despite their unusual properties.  They appear especially anomalous when the morphological properties of their component galaxies are considered.  Their identification therefore depends upon the visual inspection of large numbers of galaxies, a feat which has only recently been made possible by Galaxy Zoo, together with the Sloan Digital Sky Survey.  We present the basic properties of our cluster sample, and discuss possible formation scenarios and implications for cosmology.
\end{abstract}

\begin{keywords}
galaxies: clusters: general --- galaxies: structure --- galaxies: fundamental parameters
\vspace*{1.1cm}
\end{keywords}

\section{Introduction}

For nearly as long as it has been recognised that galaxies are stellar
systems external to our own, we have known that they are not
distributed randomly throughout space, but tend to cluster together
\citep{hubble}.  This structure is now well understood by the
amplifying influence of gravity on small scale fluctuations in the
early universe.  We are able to predict, both through simulations and
analytically, the clustering of the collisionless dark matter
component that is inferred to exist from a range of observations.  It
is an important and popular fact that the initially smooth matter
distribution collapses to form haloes, roughly spherical in shape,
though with some ellipsoidal or triaxial distortions.  These high
density haloes are joined by lower density filaments, along which
smaller haloes move, to be eventually accreted by the larger haloes,
which thus grow more massive with time \citep{bubble}.

Although the dark matter component is well understood, the behaviour
of baryonic matter is necessarily more complicated.  On large scales
it is expected to follow the dark matter, and hydrodynamical
simulations demonstrate this.  However, on small scales the density
field evolves non-linearly and the densities are such that gas physics
and feedback from collapsed baryonic objects, such as stars and black
holes, become important.  On the scale of galaxy clusters the
interplay between gas and dark matter may cause the density profiles
and shapes of haloes to vary from those predicted by models based on
dark matter alone.  In the regime of galaxies, this becomes even more
likely, as here baryons dominate the matter density.

Observationally, clusters are found to host galaxy populations quite
different to the Universal average.  Their members tend to have red
colours and suppressed or entirely absent star-formation. They also
mostly possess smooth, early type morphologies, particularly toward the
centre of a cluster.  This can partly be explained by the preference
for clusters to host the most massive galaxies, together with the
observation that more massive galaxies are more likely to be red,
passive and elliptical in any environment \citep{toil}.  However,
there remains a large population of lower-mass galaxies in clusters
whose ``red and dead'' condition is in stark contrast with the
properties of their counterparts in the field.  At higher redshifts,
this dichotomy between cluster and field galaxy populations appears to
diminish, with a growing proportion of clusters containing significant
starforming components \citep{trouble}.  At $z \ga 1$ it even appears
to reverse, with clusters hosting the most actively starforming
objects.

\vspace{0.75cm}
\section{An unusual galaxy cluster}

\begin{figure*}
\centering
\includegraphics[width=\textwidth]{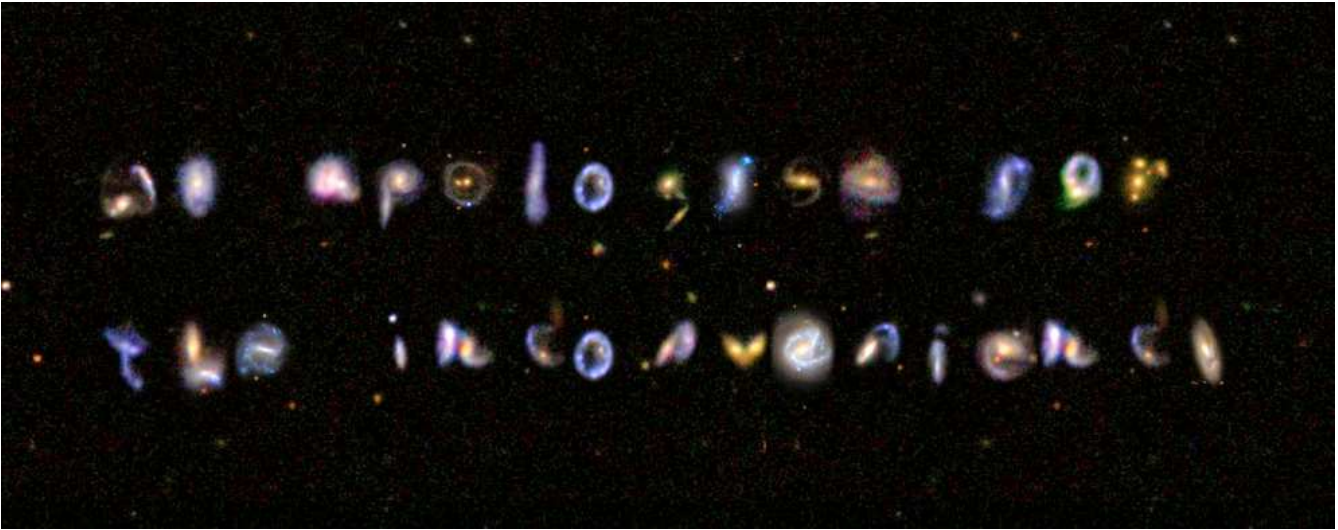}\\[0.5cm]
\caption{\label{fig1} SDSS colour composite image ($vri$) for our
  prototype unusual galaxy cluster, at $\rmn{RA}=16^{\rmn{h}}23^{\rmn{m}}76^{\rmn{s}}$, $\rmn{Dec}=+97\degr 62\arcmin 12\arcsec$, identified by Galaxy Zoo participants.  North is at the top, East is to the left.}
\end{figure*}

Given the typical properties of galaxy clusters as described above,
the existence, at low redshift ($z \sim 0.05$), of the structure
displayed in Fig.~\ref{fig1} is somewhat surprising.  Our attention
was called to this cluster by the community of Galaxy Zoo
participants, who fortuitously recognised its unusual properties
whilst classifying its individual galaxies.  The overdensity of
galaxies clearly identifies this structure as a rich galaxy cluster,
however it possesses strikingly different properties compared to
typical clusters of this richness.

One of the most distinctive aspects of this cluster is the
morphologies and colours of its component galaxies.  Many of its
members have blue colours and show clear evidence of spiral
morphology, even if the spiral arms are often disturbed.  These
disturbed morphologies are probably the result of a high frequency of
close pairs and merging systems. Such a high fraction of merging
systems is unexpected for high mass clusters due to the large velocity
dispersion, and much more typical of lower mass galaxy groups.

Another unusual aspect is the morphology of the cluster as a whole.
The structure is rather linear, and boxy, reminiscent of the filaments
seen in N-body simulations.  However, the observed galaxy density is
far higher than seen in simulations of filaments.  There is no obvious
central concentration of the number density or luminosity profile,
unlike any normal cluster of this richness.  Weak lensing and x-ray data
may assist in understanding this cluster's strange appearance, by
adding information on the distribution of the cluster's dark matter
and gas content.

Finally, but perhaps most surprising, is that upon detailed inspection,
the morphologies of individual galaxies and close systems approximate
the familiar geometric shapes of letters of the basic modern Latin
alphabet. From East to West and North to South, respectively, these
shapes may be represented as ``w e a p o l o g i s e f o r t h e i n c
o n v e n i e n c e''.  Although galaxies displaying morphologies
corresponding to Latin characters have been noticed before, `S' and `Z'
being particularly common, a localised collection of this size is
highly improbable.

A close visual inspection suggests that the galaxy distribution
exhibits an element of substructure.  The galaxies appear to divide
into five distinct groups.  These are: Group I: ``w e'', Group II: ``a
p o l o g i s e'', Group III: ``f o r'', Group IV: ``t h e'', and
Group V: ``i n c o n v e n i e n c e''.  These may be familiar to the
reader as common words of the English language.

The appearance of rational English within an astrophysical system is
widely regarded as impossible.  Furthermore, the event that an
arrangement of galaxies should express regret would be considered by
many to be ludicrous.  The data could be disregarded simply as a
statistical anomaly, an unlikely occurrence which just happens to have
occurred.  Space is, after all, not only big, but really big, and full of really
surprising things.  The authors, however, maintain that, since it is
observed, the cluster requires explanation.

It remains a possibility that previous estimates of the likelihood of
such events have been grossly underestimated and no
fundamentally new physics is required to explain this observation.
Although current cosmological simulations are not known to produce
English sentences on cluster scales, there has been little effort to
test this, and in particular a lack of visual inspection.  It is
plausible that with suitably chosen prescriptions, semi-analytic
models could reproduce an abundance of clusters similar to those
presented in this paper.

On the other hand, many would attribute a much deeper meaning to the
appearance of this cluster.  Firstly, the occurrence of these
phenomena could potentially lend support to some of the more exotic
models for Dark Energy or modified gravity, if they are able to
predict such structures.  More controversially, as most occurrences of
English sentences are considered to be the work of intelligent beings,
the existence of these messages might indicate intelligent life beyond
our own.  The scale of the messages would require a lifeform with
abilities far beyond those currently possessed by humans, and even
beyond those which we could realistically expect to acquire; implying
the existence of an intelligent being with extraordinary powers.
Indeed, another appearance of exactly the same message has been
previously reported in the hotly debated work by \citet{adams}, where
the text is interpreted as God's final message to His creation.

\vspace{0.75cm}
\section{Additional examples}

\begin{figure*}
\centering
\includegraphics[width=\textwidth]{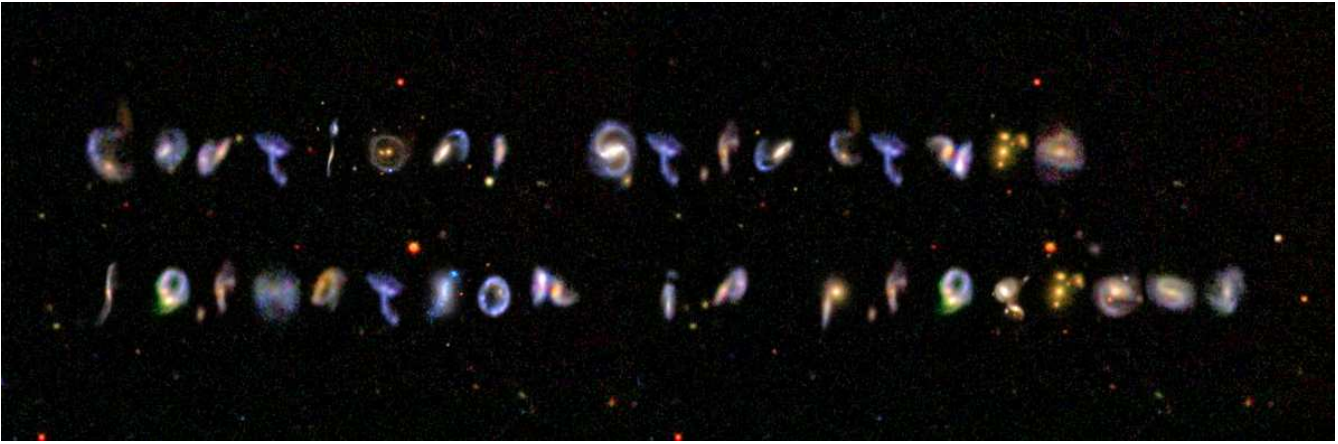}\\[0.25cm]
\caption{\label{fig2} SDSS colour composite image ($vri$) for another unusual galaxy cluster, at $\rmn{RA}=-2^{\rmn{h}}61^{\rmn{m}}12^{\rmn{s}}$, $\rmn{Dec}=+124\degr 17\arcmin 72\arcsec$, identified by Galaxy Zoo participants.  Orientation as Fig.~\ref{fig1}.}
\end{figure*}

\begin{figure}
\centering
\includegraphics[width=0.45\textwidth]{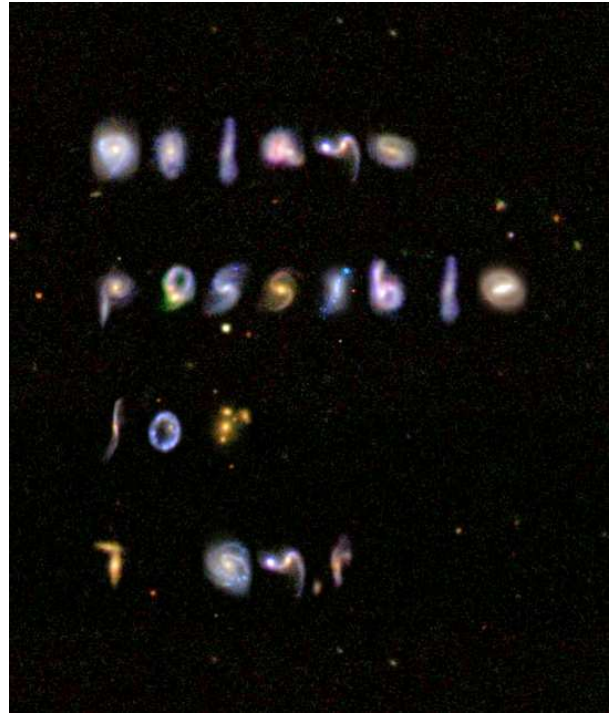}\\[0.25cm]
\caption{\label{fig3} SDSS colour composite image ($vri$) for another unusual galaxy cluster, at $\rmn{RA}=27^{\rmn{h}}10^{\rmn{m}}99^{\rmn{s}}$, $\rmn{Dec}=-97\degr 71\arcmin 23\arcsec$, identified by Galaxy Zoo participants.  Orientation as Fig.~\ref{fig1}.}
\end{figure}

The significance of the cluster discussed in the previous section is
modified somewhat by the discovery of additional examples of clusters
belonging to this unusual class.  These share many properties with the
prototype, as is clear from Figs.~\ref{fig2} \& \ref{fig3}.  In
particular, both exhibit natural subgroups of galaxies with
morphologies that conspire to resemble English words.  The cluster in
Fig.~\ref{fig2} exhibits the natural sub-structure groups ``c a u t i
o n !'', ``s t r u c t u r e'', ``f o r m a t i o n'', ``i n'', ``p r
o g r e s s'', whereas the cluster shown in Fig.~\ref{fig3} is
apparently another warning, comprising the groups ``D e l a y s'', ``p
o s s i b l e'', ``f o r'', ``7 Gyr''.

Each of the additional clusters demonstrates new features, compared
with Fig.~\ref{fig1}.  The cluster in Fig.~\ref{fig2}
appears to contain punctuation, in the form of an exclamation mark.
The cluster in Fig.~\ref{fig3}, on the other hand, includes the first
unambiguous appearance of a capital letter, ``D'', a numeral, ``7'',
and an abbreviated unit ``Gyr''.  In addition, the latter figure
demonstrates a notable left-hand justification across multiple lines.
 
Individually, these two further clusters present the same problems as
the first when considered within the context of currently
well-regarded cosmologies.  In such models, clusters that form
sensible English phrases are generally regarded as impossible.  The
three known instances, presented here, thus appear to constitute an
event that would traditionally be viewed as really not very likely at
all.  Their discovery also suggests the possibility of other messages,
not yet identified, and in particular the potential existence of a
similar clusters, utilising other languages and alphabets.

When considered collectively, the various examples presented here of
this ``unusual'' class, seem to suggest a possible common theme,
being reminiscent of the familiar local phenomenon of road works
\citep{roadworks}.  Making this identification, the message in
Fig.~\ref{fig1} may then be understood as a general
acknowledgement of blame for the specific problems conveyed in
Figs.~\ref{fig2} and \ref{fig3}.  Thus, these vivid messages are
apparently not to be understood as, in the paradigm of \citet{adams},
a message from God, but rather a notification of the common
frustrations that one group of intelligent beings imposes on other
intelligent beings in the name of progress, or even, simply, basic
maintenance of former progress.

Such a model, however, must evoke the existence of other, so called,
``intelligent beings'' beyond our own planet.  While regarded by many
to be a good long-term bet, current evidence for the existence of
extra-terrestrial life is in seriously short supply.  Even the
predictions of how much intelligent life we might reasonably expect to
find are ambiguous at best.  Indeed, some of the most rigorous
arguments on the subject actually find in favour of a total absence of
intelligent life of any kind \citep{universe}.  A suitably advanced
civilisation capable of fashioning galactic sized structures into
directed notifications, therefore, tends towards the absurd.  From
this vantage, we cannot exclude the alternative that the appearance of
familiar English phrases of unified sense in large scale cluster
morphologies are anything more than chance occurrences, which one
might hope to better understand via future insights into
probability theory or cosmology.
  
If we interpret these unusual clusters in this manner, we must
necessarily re-evaluate our understanding of their local counterparts
\citep{roadworks}.  Observations that hitherto had been taken as
certain indications the presence of intelligent life are then reduced
to nothing more than the product of pure chance.

\section{Conclusions}

Thanks to the visual inspection of SDSS images afforded by the Galaxy
Zoo project, we have identified a new class of galaxy clusters which
possess number of unusual properties.  These clusters are unusually
elongated, possess young and highly dynamic galaxy populations, and
most unexpectedly, present neatly typeset, left-justified, messages
written in the English language.  One interpretation for the existence
of these galaxy clusters is as conclusive evidence for
intelligent life elsewhere in the universe.  Conversely, however, they
could indicate that many phenomena usually attributed to intelligent
life on Earth actually occur spontaneously, without any thought
necessarily being involved at all.

\section*{Acknowledgements}

This work has been made possible by the participation of many members
of the public in visually classifying SDSS galaxies on the Galaxy Zoo
website.  Their contributions, many individually acknowledged at
http://www.galaxyzoo.org/Volunteers.aspx, have produced a number of
published scientific papers, with many more yet to come.  This article
is particularly indebted to those who have tirelessly sought out odd
and unusual objects and brought them to general attention on the
Galaxy Zoo Forum.\footnote{We stress that, despite their implausible
  appearance, the galaxies comprising each character in the figures
  presented in this paper are taken directly from the SDSS multicolour
  composite imaging.  Note, however, that some degree of translation
  and rotation has been performed to the individual characters, for
  presentation purposes.}  We thank them for their extraordinary
efforts in making this project a success. We are also grateful to
various members of the media, both traditional and online, for helping
to bring this project to the public's attention.

Funding for the Sloan Digital Sky Survey (SDSS) and SDSS-II has been
provided by the Alfred P. Sloan Foundation, the Participating
Institutions, the National Science Foundation, the U.S. Department of
Energy, the National Aeronautics and Space Administration, the
Japanese Monbukagakusho, and the Max Planck Society, and the Higher
Education Funding Council for England. The SDSS Web site is
http://www.sdss.org/.

The SDSS is managed by the Astrophysical Research Consortium (ARC) for
the Participating Institutions. The Participating Institutions are the
American Museum of Natural History, Astrophysical Institute Potsdam,
University of Basel, University of Cambridge, Case Western Reserve
University, The University of Chicago, Drexel University, Fermilab,
the Institute for Advanced Study, the Japan Participation Group, The
Johns Hopkins University, the Joint Institute for Nuclear
Astrophysics, the Kavli Institute for Particle Astrophysics and
Cosmology, the Korean Scientist Group, the Chinese Academy of Sciences
(LAMOST), Los Alamos National Laboratory, the Max-Planck-Institute for
Astronomy (MPIA), the Max-Planck-Institute for Astrophysics (MPA), New
Mexico State University, Ohio State University, University of
Pittsburgh, University of Portsmouth, Princeton University, the United
States Naval Observatory, and the University of Washington.

\bsp

\label{lastpage}

\end{document}